\documentclass[]{elsart}
\usepackage{epsfig}
\input axodraw.sty
\newcommand{\mett}{{\not\!\!E}_{T}}

\def\beq{\begin{equation}}
\def\eeq{\end{equation}}
\def\bea{\begin{eqnarray}}
\def\eea{\end{eqnarray}}

\newcommand{\threegraphs}[3]{%
\unitlength=1in
\begin{picture}(5.4,5)
\put(1.35,0){\epsfig{file=#3.eps, width=2.7in}}
\put(0,2.5){\epsfig{file=#1.eps, width=2.7in}}
\put(2.7,2.5){\epsfig{file=#2.eps, width=2.7in}}
\put(1.35,2.3){(c)}
\put(0,4.8){(a)}
\put(2.7,4.8){(b)}
\end{picture}}

\journal{Physics Letters}

\begin{document}
\hfill\parbox{8cm}{\raggedleft CERN-TH/2002-255 \\ LAPTH-942/02 \\
  TIFR/TH/02-31 \\ hep-ph/0210375 \\
}
\begin{frontmatter}
\title{Combined fits to the supersymmetric explanation of anomalous
  lepton-$\gamma$-missing $E_T$ events}

\author[LAPP]{B.C.  Allanach}, \author[tata]{K. Sridhar}

\address[LAPP]{LAPTH, Chemin de Bellevue, B.P. 110,
  Annecy-le-Vieux 74951, France}
\address[tata]{Department of Theoretical Physics, Tata Institute of
  Fundamental  
Research, Homi Bhabha Road, Mumbai 400 005, India}

\begin{abstract}The CDF experiment reported a lepton photon missing transverse
  energy ($\mett$) signal 3$\sigma$ in excess of the
  Standard Model prediction in Tevatron Run I data. The excess can be
  explained by the resonant production of a smuon, which subsequently decays
  to a muon, a photon and a gravitino. Here, we perform combined fits of this
  model to the CDF $\gamma l \mett$ excess, the D0 measurement of the same
  channel and
  the CDF $\gamma \mett$ channel.
  Although the rates of the latter two analyses are in agreement with the
  Standard Model prediction,
  our model is in good agreement with these data because their signal to
  background efficiency is low at the best-fit point. 
  However, they help to constrain the model away from
  the best fit point.
\end{abstract}
\end{frontmatter}

\section{Introduction}
The CDF experiment has recently discovered an anomaly in the production 
rate of lepton-photon-$\mett$ in $p{\bar p}$ collisions at $\sqrt{s}= 1.8 
~{\rm TeV}$, using 86.34 pb$^{-1}$ of Tevatron 1994-95 data \cite{CDF}.
While the number of events expected from the Standard Model (SM) were
$7.6\pm 0.7$, the experimentally measured number corresponded to 16. 
Moreover, 11 of these events involved muons (with 4.2 $\pm$ 0.5 expected)
and 5 electrons (with 3.4 $\pm$ 0.3 expected). 

\begin{figure}
\begin{picture}(400,170)
\ArrowLine(60,10)(120,50)
\ArrowLine(120,50)(60,90)
\DashLine(120,50)(180,50){5}
\ArrowLine(180,50)(240,10)
\DashLine(180,50)(240,90){5}
\DashLine(240,90)(300,70){5}
\Photon(240,90)(300,130){5}{4}
\put(65,5){$q$}
\put(65,90){$\bar q'$}
\put(245,10){$\mu$}
\put(210,80){$\chi_0$}
\put(150,65){$\tilde\mu$}
\put(305,65){$\tilde G$}
\put(305,125){$\gamma$}
\end{picture}
\caption{Feynman diagram of resonant smuon production followed by 
neutralino decay.}
\label{feyn}
\end{figure}
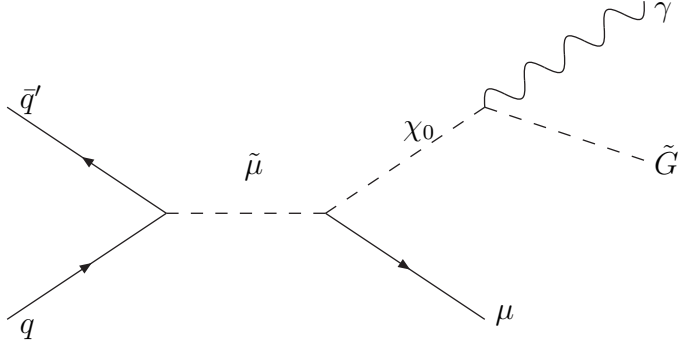

In earlier papers \cite{us} we suggested that the excess can be simply 
understood in terms of the minimal supersymmetric standard model (MSSM) which
has the following 
ingredients: (1) the model is $R$-parity violating with an $L$-violating 
$\lambda'_{211}$ coupling, and (2) the supersymmetric spectrum includes
an ultra-light gravitino of mass $\sim10^{-3}$ eV. We
have demonstrated in our earlier papers \cite{us} that such a model
provides a natural explanation of the CDF anomaly and explains not
only the excess in the number of $\mu \gamma \mett$ events but also
explains the main features of the kinematic distributions of the excess
events. The excess can be explained using a small value of the $L$-violating 
coupling $\lambda'_{211}\sim 0.01$ because of the resonant production of
smuons in  the annihilation of an initial-state $q \bar q$ pair. 
The smuon thus produced 
decays predominantly into a bino-dominated neutralino and a muon, with the 
neutralino further decaying into a photon and a gravitino. The production 
and decay has been shown in the Feynman diagram in Fig.~\ref{feyn}. 
The 
fact that the excess is seen in final states involving photons emerges 
very neatly in our model because the decay $\chi_1^0 \rightarrow \gamma 
\tilde G$ dominates overwhelmingly over other decay modes. 

In the current article, we extend the previous studies by including two
additional pieces of independent empirical
information. We include the D0 Run I measurement~\cite{D0} of the 
$\mu \gamma$ missing
$E_T$ process, as well as $\gamma$ missing $E_T$ data coming from 
CDF in Run I~\cite{CDF2}. The empirical background event-rate in the $\mu
\gamma \mett$ channel is  
quite different in the CDF and D0 cases due to the different cuts employed.
Our scenario predicts excesses in each of these channels, and we
determine to what extent it is in accord with their measured rates.
By performing a combined fit to all three event rates, we constrain the masses
of the relevant sparticles in the event, as well as $\lambda'_{211}$.

\section{The model}
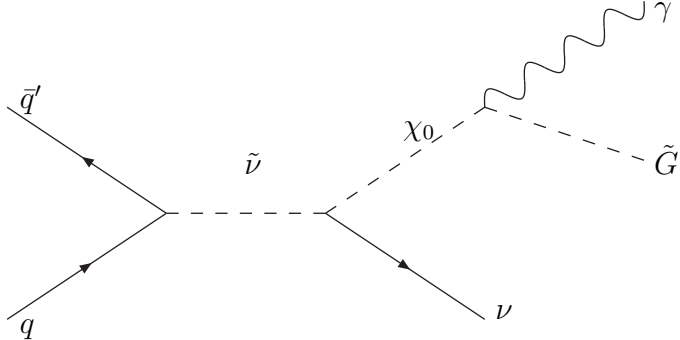
\begin{figure}
\begin{picture}(400,170)
\ArrowLine(60,10)(120,50)
\ArrowLine(120,50)(60,90)
\DashLine(120,50)(180,50){5}
\ArrowLine(180,50)(240,10)
\DashLine(180,50)(240,90){5}
\DashLine(240,90)(300,70){5}
\Photon(240,90)(300,130){5}{4}
\put(65,5){$q$}
\put(65,90){$\bar q'$}
\put(245,10){$\nu$}
\put(210,80){$\chi_0$}
\put(150,65){$\tilde\nu$}
\put(305,65){$\tilde G$}
\put(305,125){$\gamma$}
\end{picture}
\caption{Feynman diagram of resonant muon sneutrino production followed by 
neutralino decay, resulting in a $\gamma \mett$ signature.}
\label{feyn2}
\end{figure}

In order that the cross-section for the production of the smuon resonance
is substantial enough to account for the anomalous events, we need the 
left-handed
smuon to be light i.e. $m_{{\tilde \mu}_L}\sim150$ GeV. Further, one needs 
to couple the smuon to valence quarks in the initial state
implying that the $L$-violating operator that we need is $L_2Q_1\bar{D}_1$ 
corresponding to the coupling $\lambda'_{211}$, which generates the 
interactions ${\tilde \mu}u\bar{d}$ and ${\tilde \nu}_\mu d \bar{d}$ 
(and charge conjugates), along with other supersymmetrised copies involving 
squarks. Therefore, the operator we invoke in our model 
predicts supersymmetric signals in other channels which manifest themselves
through the production of either sneutrinos or squarks. In our model, 
we take the squarks to be heavy and so their effects on experimental 
observables will be negligible. On the other hand, the spontaneously broken
SU(2)$_L$ symmetry in the MSSM implies that muon sneutrinos have a
tree-level mass squared~\cite{softsusy}
\begin{equation}
m_{\tilde \nu}^2 = m_{{\tilde \mu}_L}^2 - m_\mu^2 + (1 - \sin^2 \theta_w) \cos
(2 \beta) M_Z^2 \qquad \Rightarrow \qquad m_{\tilde \nu} < m_{{\tilde \mu}_L},
\end{equation}
where $m_\mu$ is the mass of the muon, $\theta_w$ the Weinberg angle, and $\tan
\beta$ the ratio of the MSSM Higgs vacuum expectation values ($\cos 2
\beta < 0$). Since $m_{\tilde \nu} < m_{{\tilde \mu}_L}$ 
if we assumed that smuons were produced
at the Tevatron Run I energy, we should expect that muon sneutrinos could also
be produced. 
The dominant production mechanism is resonant sneutrino production and
subsequent decay, as shown in Fig.~\ref{feyn2}. It results in a $\gamma \mett$
experimental signature.

The pattern of masses of the super-particles suggested by the CDF $\mu \gamma
\mett$ anomaly
is the following: the smuon is around
150 GeV, and the only other light sparticles are the neutralino (which
is typically about 45 GeV lighter than the smuon) and the 
ultra-light gravitino (which is as light as $10^{-3}$ eV). We enforce
degeneracy between the first two generations in order to avoid flavour
changing neutral currents. Other sparticles do not play a role in this
analysis, and are set to be arbitrarily heavy.
Such a light
gravitino materialises naturally in models of gauge-mediated supersymmetry
breaking (GMSB)\cite{gmsb}. However, in the minimal version of GMSB models
the chargino and the second-lightest neutralino are not much heavier than 
the neutralino and this feature of the minimal model is certainly not 
desirable for our considerations because it leads to large jets$+ \gamma +
\mett$ rates which are not seen by experiments\footnote{This 
observation has been made earlier in the literature \cite{baer}
in the context of the GMSB-based explanation \cite{kane} of the
$ee \gamma \gamma \mett$ CDF event~\cite{abe}.}. If the minimal versions
of GMSB models do not yield the pattern of super-particle masses that
we need, then the obvious question to ask is what is the high-energy
model that yields this mass spectrum at low energies. It is interesting
to note that such a mass spectrum can arise in an alternate model of GMSB 
which is obtained from compactifying 11-dimensional M-theory on a 7-manifold 
of G2 holonomy~\cite{witten}. For the purposes of this paper, however, we
simply work with the low-energy model with the mass spectrum described
above and do not worry about the high energy completion of this model.

\section{Simulating the experiments}

In our model, we have essentially four free parameters that are relevant to
the data we fit: the
gravitino mass, $m_{\tilde G}$, the neutralino mass $m_{\chi_1^0}$, the smuon
mass, $M_{\tilde \mu}$ and the $R$-violating coupling 
$\lambda'_{211}$. However, instead of simultaneously fitting the four 
parameters using the experimental data, we choose to work with fixed values 
of two of these parameters close to their best-fit values while performing
fits in the two other parameters.
For our analysis, we take other parameters like ${\rm tan} \beta=10$ to
be constant. The coupling, $\lambda'_{211}$, is constrained from  
$R_\pi = \Gamma (\pi \rightarrow e \nu) / (\pi \rightarrow \mu \nu)$
\cite{bgh} to be $< 0.059 \times \frac{m_{\tilde{d_R}}}{100~{\rm GeV}}$ 
\cite{rparrev}. But since the constraint involves a squark mass which is 
large in our model, it is not very relevant. While the production of the 
smuon resonance is through the $R$-violating mode, to produce the $l \gamma
\mett$ excess we require that its
decay goes through the 
$R$-conserving channel to a neutralino and muon final state. The $R$-violating
decay of the slepton is possible but constrained, in principle, by the
Tevatron di-jet data \cite{cdfjets} which exclude a $\sigma . B> 1.3 \times 
10^4$ pb at 95\%~C.L. for a resonance mass of 200 GeV. However, in practice
this does not provide a restrictive bound upon our scenario as long as 
the $R$-violating coupling is sufficiently small, $< \mathcal{O}(1)$. We also
add that the di-jet
bound is not restrictive because it suffers from a huge QCD background.
By restricting $\lambda'_{211}$ to be small, we also avoid significant rates
for the possible 
$R$-violating decays of $\chi_1^0 \rightarrow \mu jj$ or 
$\chi_1^0 \rightarrow \nu jj$.

%

We use the {\small \tt ISASUSY} part of the {\small \tt ISAJET7.58}
package~\cite{isajet} to generate the spectrum, branching ratios and 
decays of the sparticles. For an example of parameters, we choose 
(in the notation used by ref.~\cite{isajet}) $\tan \beta=10$ and
$A_{t,\tau,b}=0$.
$\mu$ together with other flavour diagonal soft supersymmetry breaking 
parameters are set to be very large. We emphasise that this is a 
representative point in the supersymmetric parameter space and not a 
special choice. 

As stated earlier, we present analyses for three sets of data in this paper:
the CDF Run I data on $l \gamma \mett$~\cite{CDF}, the D0 Run I data on 
$\mu \gamma \mett$~\cite{D0} and the CDF Run I data on 
$\gamma \mett$~\cite{CDF2}. 
We now present our fiducial efficiencies and cuts, which mimic those of the
relevant experiments.
The CDF 
experiment detects photons with the constraints that 
the following pseudo-rapidity regions are excluded: $|\eta^{\gamma}|>1$, 
$|\eta^{\gamma}|<0.05$ and the region $0.77<\eta^{\gamma}<1.0, 
75^\circ<\phi<90^\circ$. For the $l \gamma \mett$ data, photon detection
efficiency within these cuts is $\eta_\gamma=81\%$.
Muons have a 60$\%$ detection efficiency if $|\eta_\mu|<0.6$ or 45$\%$
if $0.6\leq \eta_\mu\leq 1.1$. To improve signal to background ratio
efficiency, the following
cuts are used: $E_T(\mu)>25$ GeV, $E_T(\gamma)>25$ GeV and $\mett>25$ GeV.
\begin{table}
\begin{center}
\begin{tabular}{c|cccc}
$E_T(\gamma)$/GeV & $<$60 & 60-65 & 65-80 & $>$80 \\ \hline
efficiency        & 40\%  & 47\%  & 51\%  & 54\%  \\
\end{tabular}
\end{center}
\caption{Photon efficiency in the CDF $\gamma \mett$ analysis}
\label{effCD}
\end{table}
For the $\gamma \mett$ analysis, 
the CDF experiment~\cite{CDF2} has chosen the following cuts: 
$\vert \eta^{\gamma} \vert \le 1.0$, $E_T^{\gamma} >$ 55 GeV and
$\mett >$ 45 GeV. The cut on the photon $E_T$ is chosen to be as large
as 55 GeV so as to beat down the background due to cosmic rays.
Photon fiducial efficiencies are shown in table~\ref{effCD}, 
and were
obtained from ref.~\cite{CDF2}
 
In contrast, in order to simulate the D0 experiment, we specify
$\vert \eta_{\mu} \vert \le 1.0$,
$\vert \eta_{\gamma} \vert \le 1.1$ or $1.5 \le \vert \eta_{\gamma} 
\vert \le 2.5$.
We use the same cuts as D0: $p_T^\mu > 15$ GeV,
$p_T^\gamma >
10$ GeV on the muon and photon $p_T$ respectively.  Also, $\mett \ge$ 15 GeV,
$\Delta R (\mu \gamma) \ge 0.7$
and $M_T(\mu \mett)\ge 30$ GeV.  
Here, $\Delta R(\mu \gamma)$ is the distance between the photon and muon in
pseudo-rapidity ($\eta$) 
and transverse angle ($\phi$) space. $M_T^2 \equiv E_T^2 - p_T^2$ is the
transverse mass. Within these cuts, we have fiducial efficiencies of
$71.1 \%$, $50.1 \%$ and $51.0\%$ for the trigger, muon and photon
respectively. 
These cuts have been optimised by D0 with a view to
studying the effects of anomalous gauge boson couplings on this final
state. Unfortunately, the signal to background ratio for our model in the D0
analysis is far from optimal for
the signal that we propose to study. 

%

\begin{table}
\begin{center}
\begin{tabular}{c|c|c|c}
experiment & luminosity & observed number & background \\ \hline
CDF $l \gamma \mett$ &86 pb$^{-1}$&  11 & $4.2 \pm 0.5$ \\
D0  $\mu \gamma \mett$ &73 pb$^{-1}$& 58 & $58 \pm 9.75$ \\
CDF $\gamma \mett$     &86 pb$^{-1}$& 11 & $11 \pm 2.2$ \\
\end{tabular}
\end{center}
\caption{Observed number of events passing cuts in the text and Standard
  Model backgrounds for the 
  three pieces of data included in the combined fit. The integrated
  luminosity for each analysis is also listed.}
\label{expnums}
\end{table}
The observed number of events in the different analyses, and their 
Standard Model backgrounds (taken from the experimental papers~\cite{CDF,D0,CDF2}) are shown in Table~\ref{expnums}. 

We now simulate the signal events for the processes in
Figs.~\ref{feyn},\ref{feyn2}.
We use {\small \tt HERWIG6.4}~\cite{herwig} including 
parton showering (but not including jet isolation cuts) to calculate 
cross-sections for single slepton production. 

\section{Combined fits}

For each of the three data listed in Table~\ref{expnums}, 
we can define a log
likelihood defined by the Poissonian log-likelihood for $n_o$ observed
events, $n_s$ expected signal events (for fixed values of SUSY breaking
parameters), convoluted with a Gaussian probability distribution of the number
of expected Standard Model background events $n_{SM}$ and its uncertainty
$\sigma_{SM}$: 
\begin{equation}
\ln \mathcal{L}(n_s) \equiv \frac{1}{\sqrt{2 \pi} \sigma_{SM}}  \int_0^\infty  
e^{\frac{-(n - n_{SM})^2}{2 \sigma_{SM}^2}} 
\left( n_o \ln (n + n_S) - (n+ n_s) - \ln n_o \right) dn.
\end{equation}
This is a good approximation provided $n_{SM}$ is several times $\sigma_{SM}$
above zero, as is the case here.
$\ln (\mathcal{L}(n_s) / \mathcal{L}(0))$ is then the (signal+background) to
background likelihood ratio for a given analysis. We can form the total fit
likelihood ratio $\ln\mathcal{L}^{tot}$ by adding the likelihood ratio from
each of the three analyses in Table~\ref{expnums}.
We will always consider two
relevant 
parameters to be fixed and fit the model to the other two. The best-fit point
in parameter space corresponds to the maximum value of the likelihood ratio
and is found by using {\small \tt MINUIT}~\cite{minuit}.
$-2 \ln \mathcal{L}^{tot}_{max}$ corresponds to the equivalent number of
$\chi^2$(Standard Model) - $\chi^2$(best-fit)$\equiv \Delta \chi^2$.
We have one degree of freedom, and therefore the 90$\%$ and 95$\%$ confidence
level ($C.L.$) limits on parameters then lead to~\cite{minuit} 
$\ln \mathcal{L}^{tot}_{max} - \ln \mathcal{L}^{tot} = 1.35,1.92$ respectively
in the fit. We will constrain $M_{\chi_1^0}>100$ GeV, as implied by 
$\gamma \gamma \mett$ LEP2 data~\cite{lep2}.
We now discuss the results of combined fits for various
hyper planes of parameter space.

\begin{figure}
\threegraphs{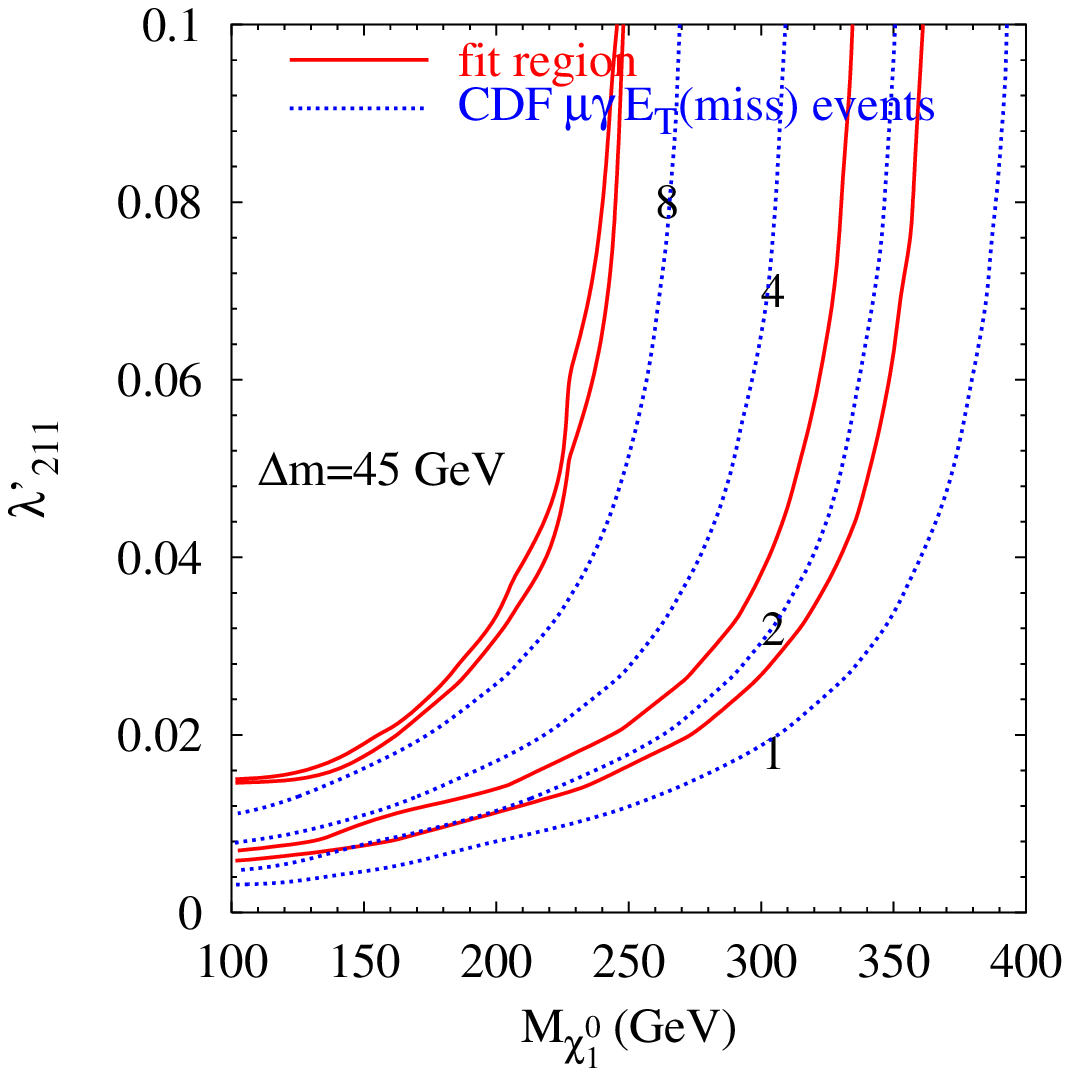}{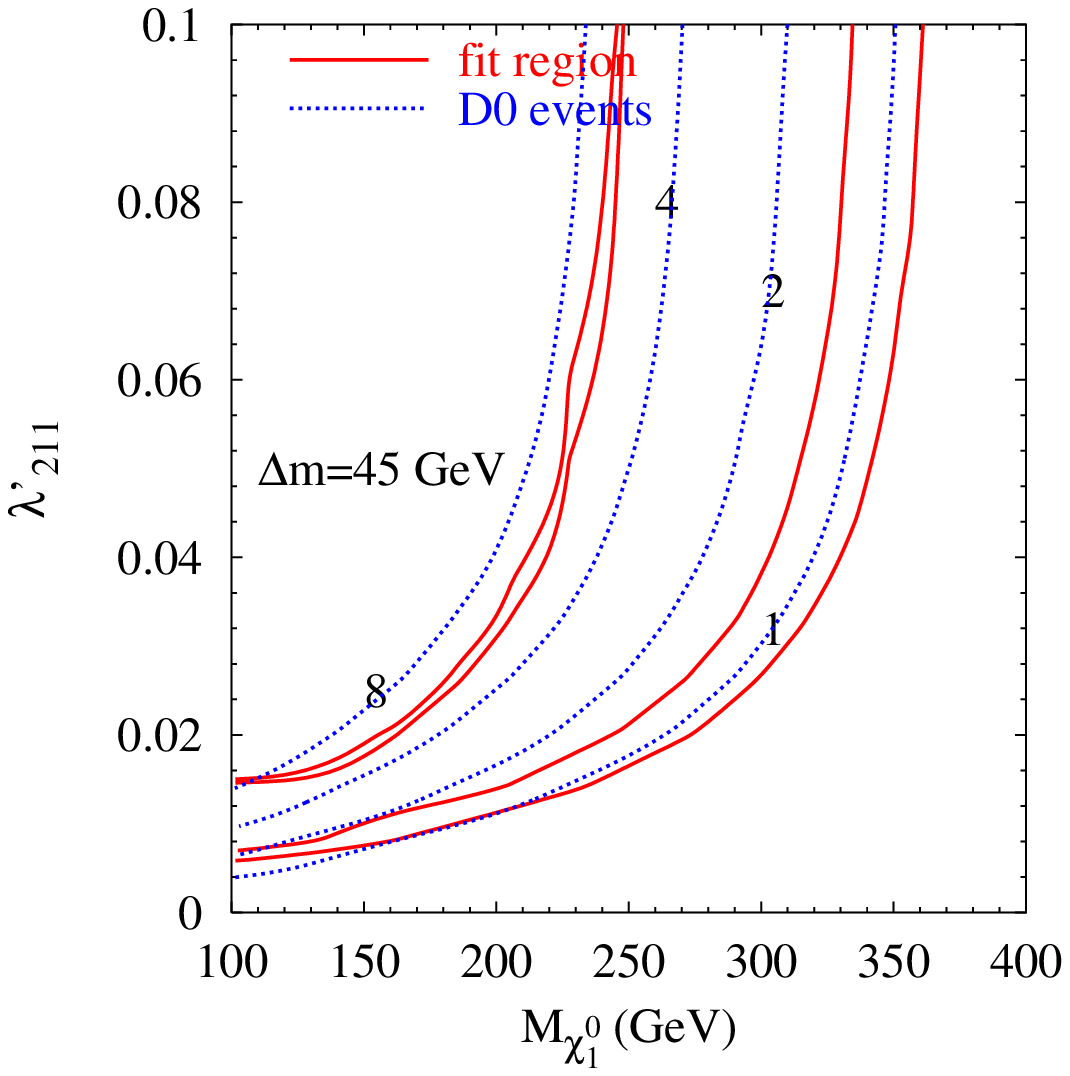}{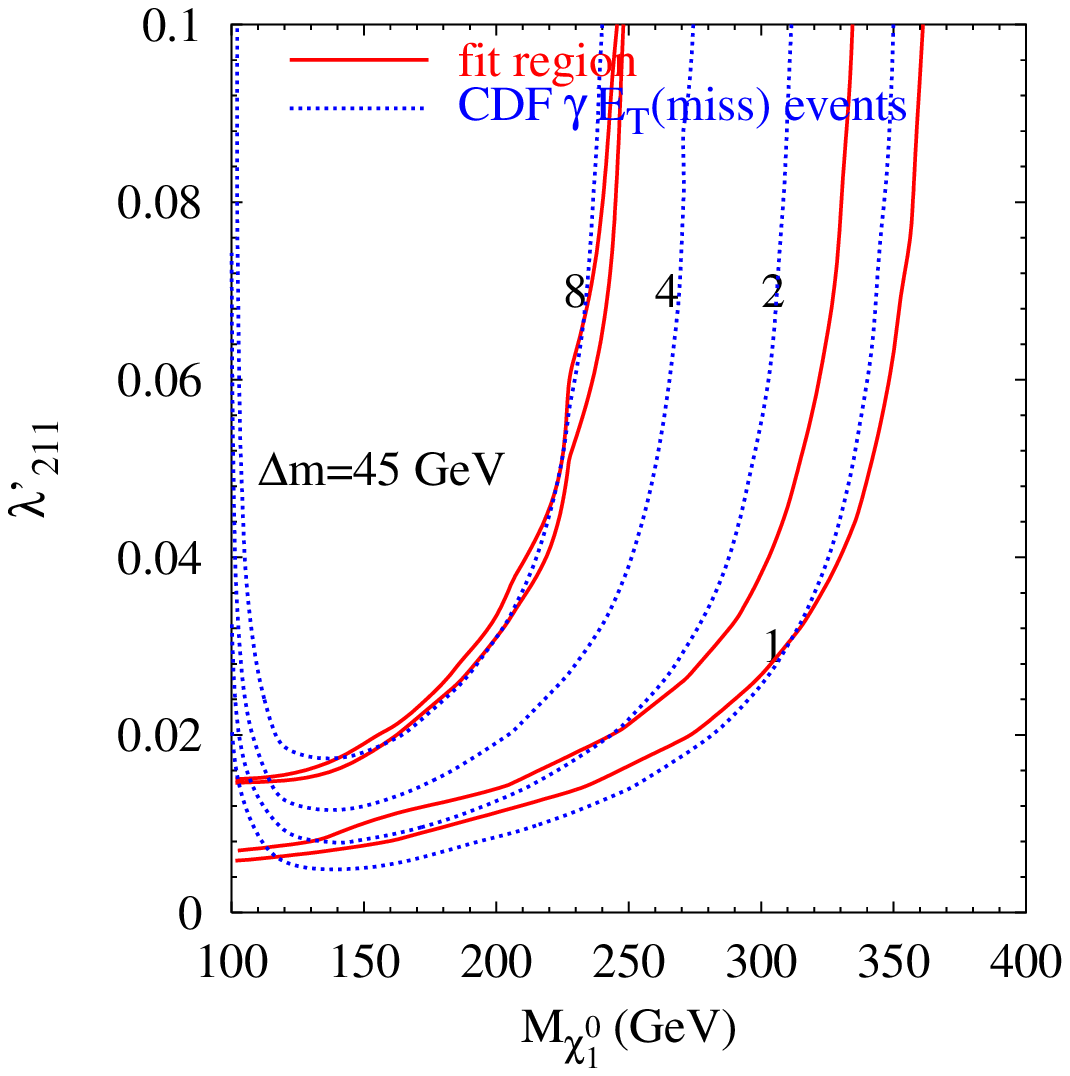}
 \caption{Predicted number of excess events in (a) CDF $l \protect\gamma \protect\mett$,
   (b) D0 $l \protect\gamma \protect\mett$,
   and (c) CDF $\gamma \mett$  channels in the $\protect\lambda'_{211}$-neutralino
   mass plane for $\tan \beta=10$, $\Delta m=45$ GeV and $m_{\tilde
   G}=10^{-3}$. Labelled contours of equal numbers of signal events are shown
   as    dashed curves.
   The 90$\protect\%~C.L.$ and 95$\protect\%~C.L.$ combined-fit regions are
   between the inner and outer pair of solid lines respectively.}
\label{lamm1}
\end{figure}
Fig.~\ref{lamm1} 
displays the $90$ and $95\%~C.L.$ fit regions as the area
between the solid lines in the $\lambda'_{211}, M_{\chi_1^0}$ plane. 
$\Delta m\equiv M_{\tilde \mu} - M_{\chi_1^0}$ has been kept fixed at 45 GeV
in order to keep the decay mode ${\tilde \mu} \rightarrow \mu \chi_1^0$ open 
and $m_{\tilde G}=10^{-3}$ eV. A significant amount of parameter space fits
the combined data, with ranges $\lambda'_{211}>0.001$. Increasing $M_{\chi_1^0}$ produces a lower cross-section because of 
kinematical suppression, but this effect can be compensated by raising
$\lambda'_{211}$, thus increasing the production
rate.
The
best-fit point is $\lambda'_{211}=0.11$,
$M_{\chi_1^0}=100$ GeV, with $\Delta \chi^2 = 6.90$. 
Fig.~\ref{lamm1}a shows that we expect between 3-8
signal events in the CDF $l \gamma \mett$ channel. This data dominates the
fit because backgrounds (and their uncertainties) are larger in the other
analyses. 1.5-6 events are 
expected at D0, and between 1.5 and 8 signal events were predicted for the CDF
$\gamma \mett$ signature, depending on the parameters. 

\begin{figure}%
\threegraphs{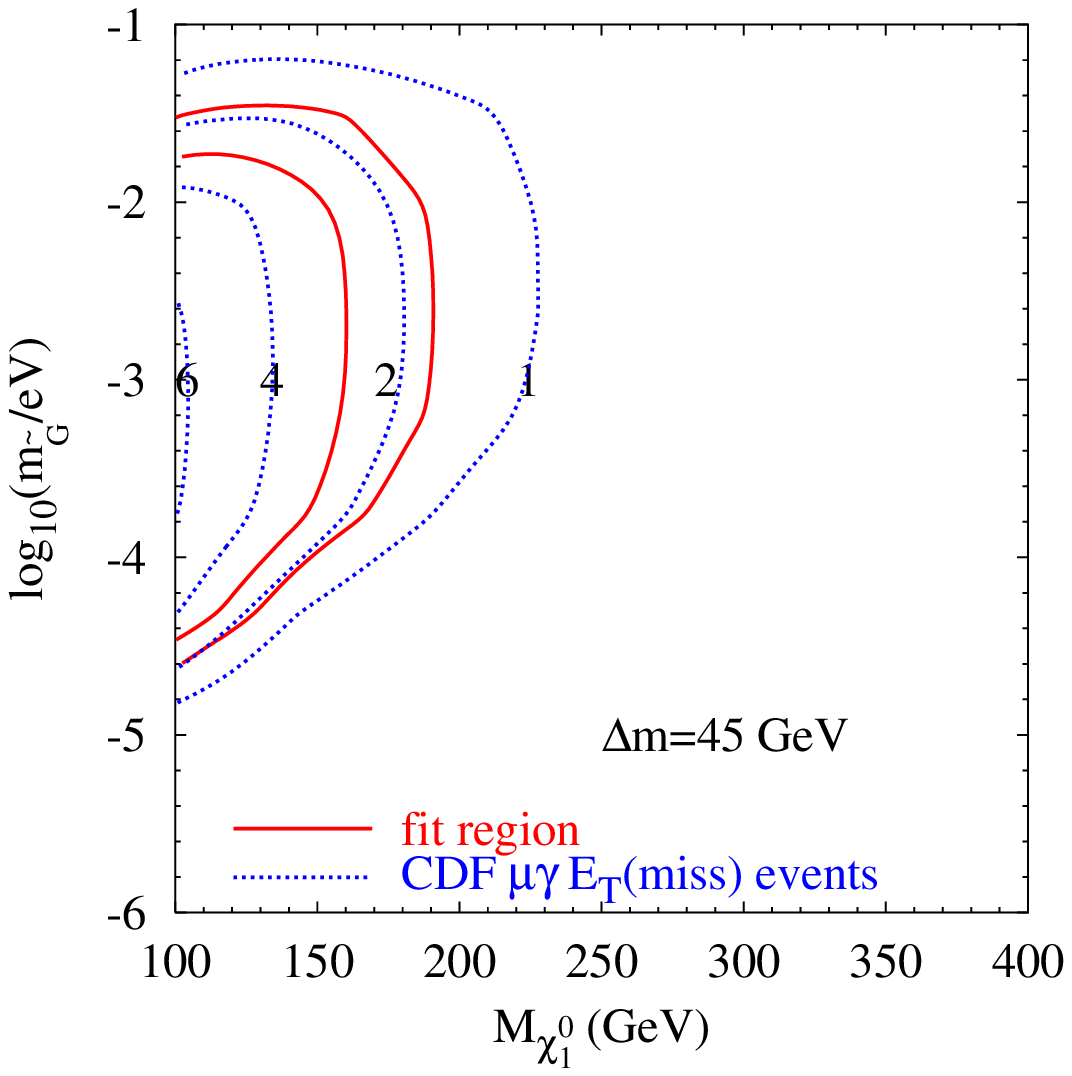}{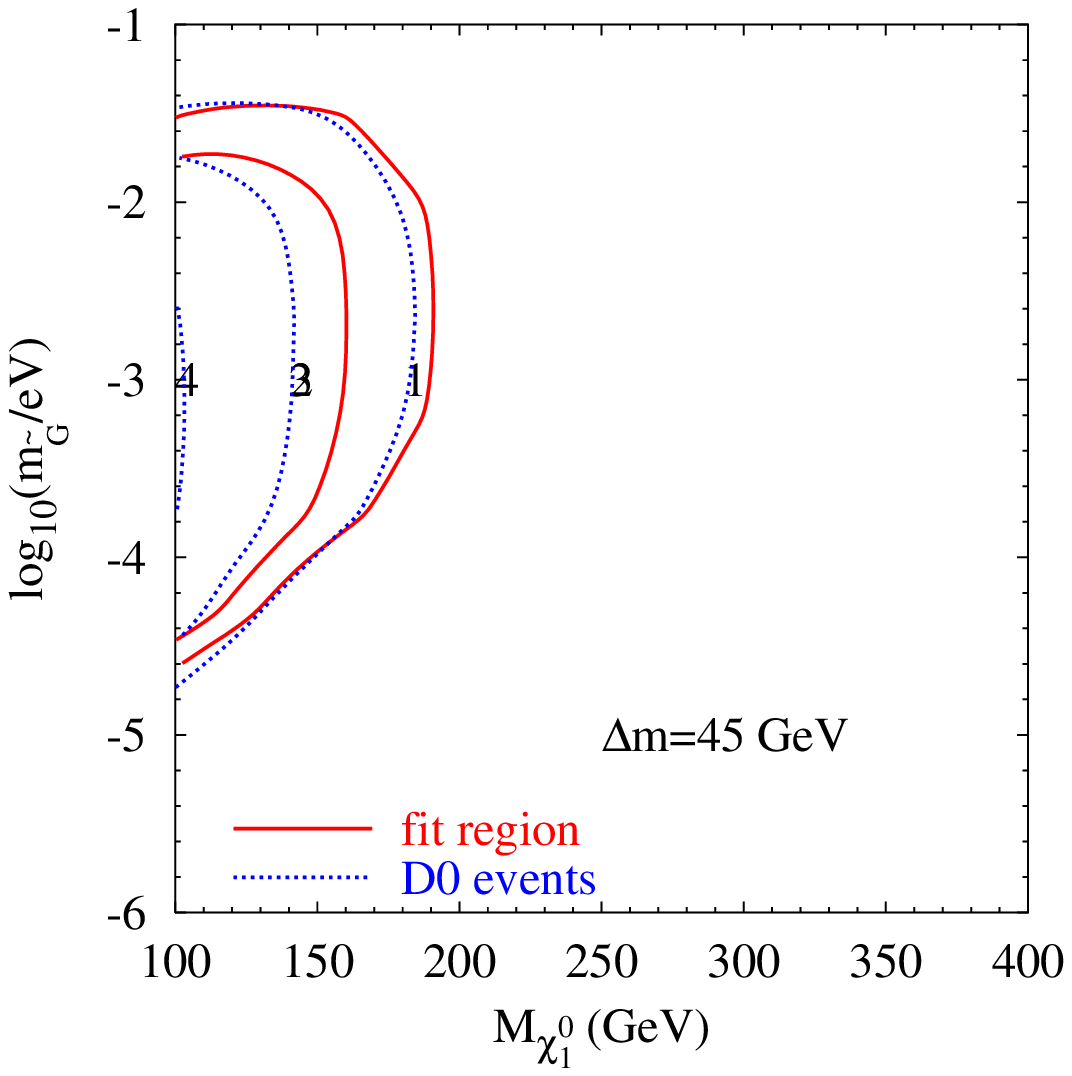}{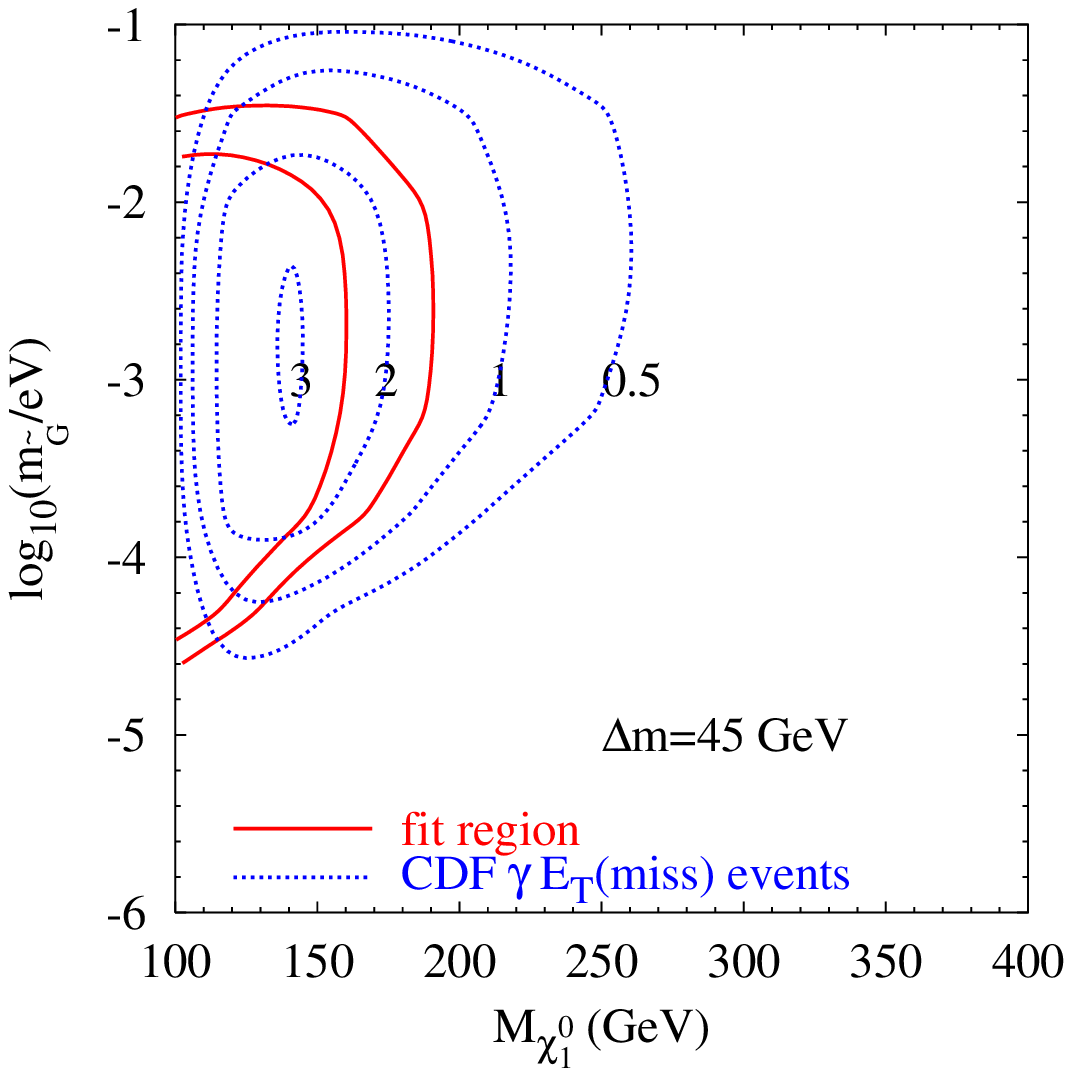}
 \caption{Predicted number of excess events in (a) CDF $l \gamma \mett$,
   (b) D0 $l \gamma \mett$,
   and (c) CDF $\gamma \mett$  channels in the gravitino mass-neutralino
   mass plane, for $\tan \beta=10, \Delta m=45$ GeV and
   $\lambda'_{211}=0.01$. Labelled contours of equal numbers of signal events
   are shown as    dashed curves.
   The 90$\%~ C.L.$ and 95$\%~C.L.$ combined-fit regions are
   to the left of the inner and outer solid lines respectively.}
\label{mgm1}
\end{figure}
Fig.~\ref{mgm1} displays the $90$ and $95\%~C.L.$ fit regions as the area
to the left-hand side of the solid lines in the $\log(m_{\tilde G}),
M_{\chi_1^0}$ plane. Here,
we use the default values $\lambda'_{211}=0.01, \Delta m=45$ GeV. 
It is clear from the fit regions that the data prefer lower values of the
neutralino mass. The best-fit
point is: $M_{\chi^1_0}=100$ GeV,
$m_{\tilde G}=10^{-3.0}$ eV, $\Delta \chi^2=6.76$. 
The figure illustrates that if $m_{\tilde G}>0.1$ eV, the branching ratio of
$\chi_1^0
\rightarrow {\tilde G} \gamma$ becomes tiny, decreasing the cross-section for 
the CDF $l \gamma \mett$ signal, which dominates the fit. When $m_{\tilde G}$
is below $10^{-5}$ eV, ${\tilde \mu} \rightarrow \mu {\tilde G}$ decays
dominate, again decreasing the CDF $\mu \gamma \mett$ signal.
2-6 CDF $l \gamma \mett$ excess events, 2-4 D0 excess events and
0-3 $\gamma \mett$ CDF excess events are expected, shown by the dashed contours
within the 90$\%~C.L.$ regions of Figs.~\ref{mgm1}(a-c).

\begin{figure}%
\threegraphs{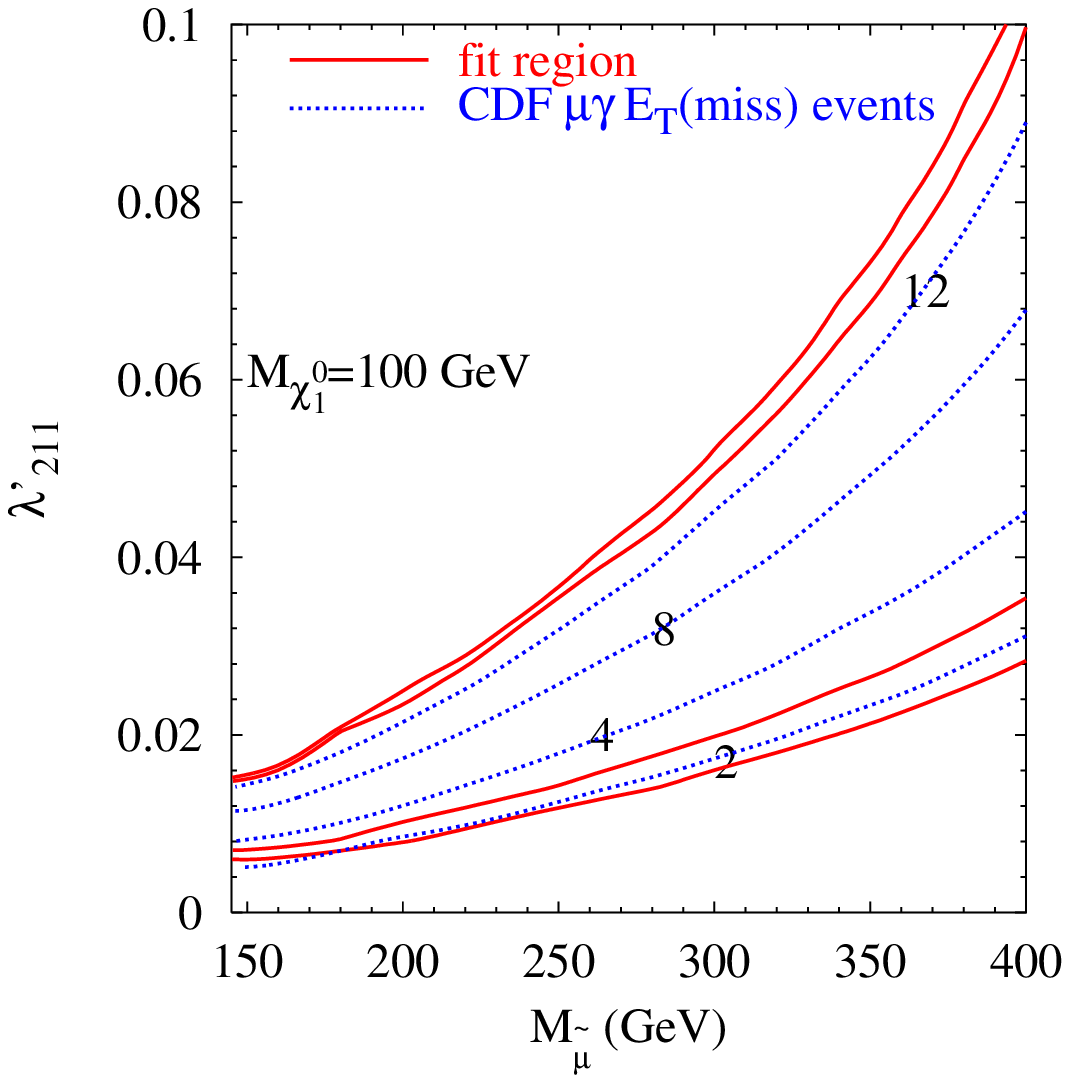}{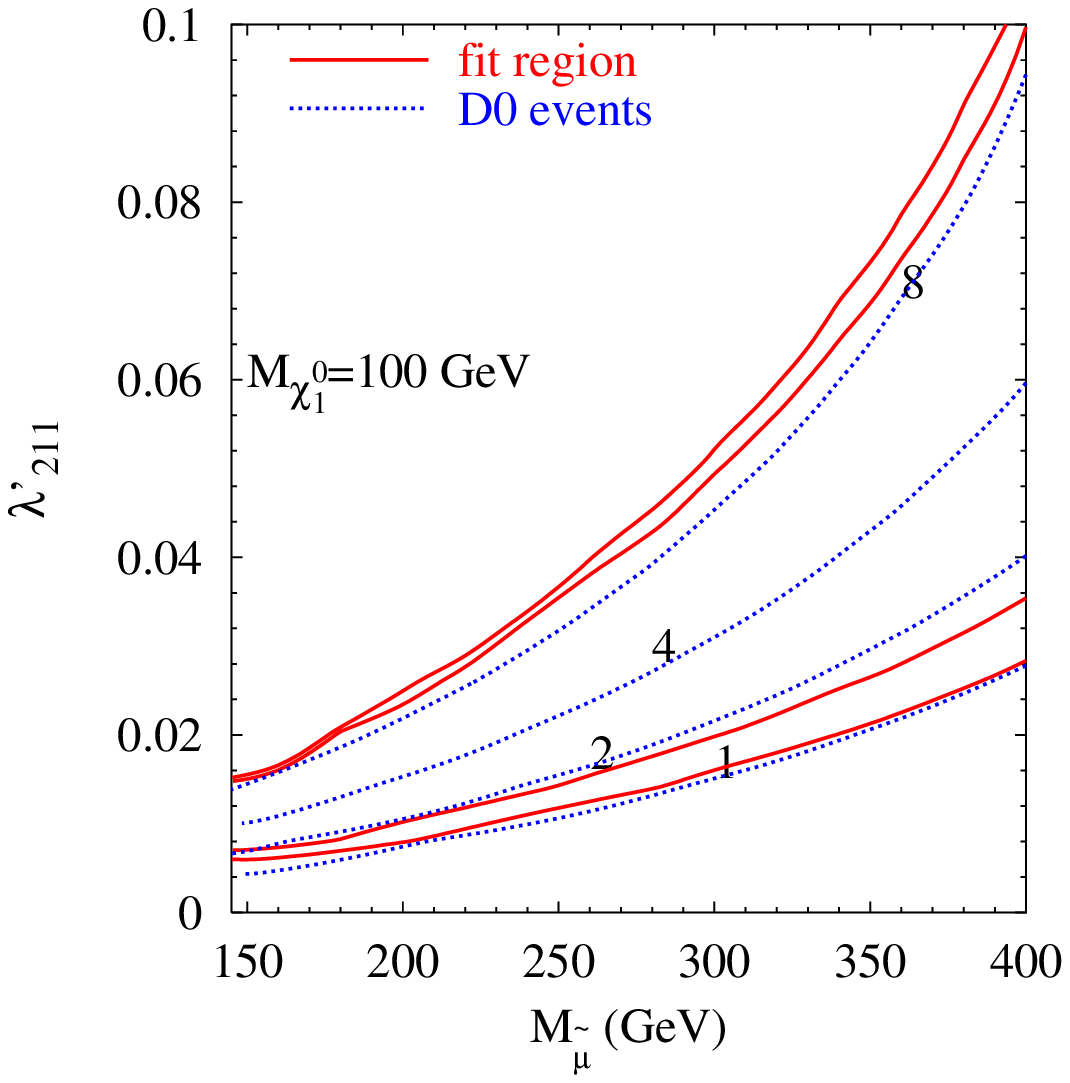}{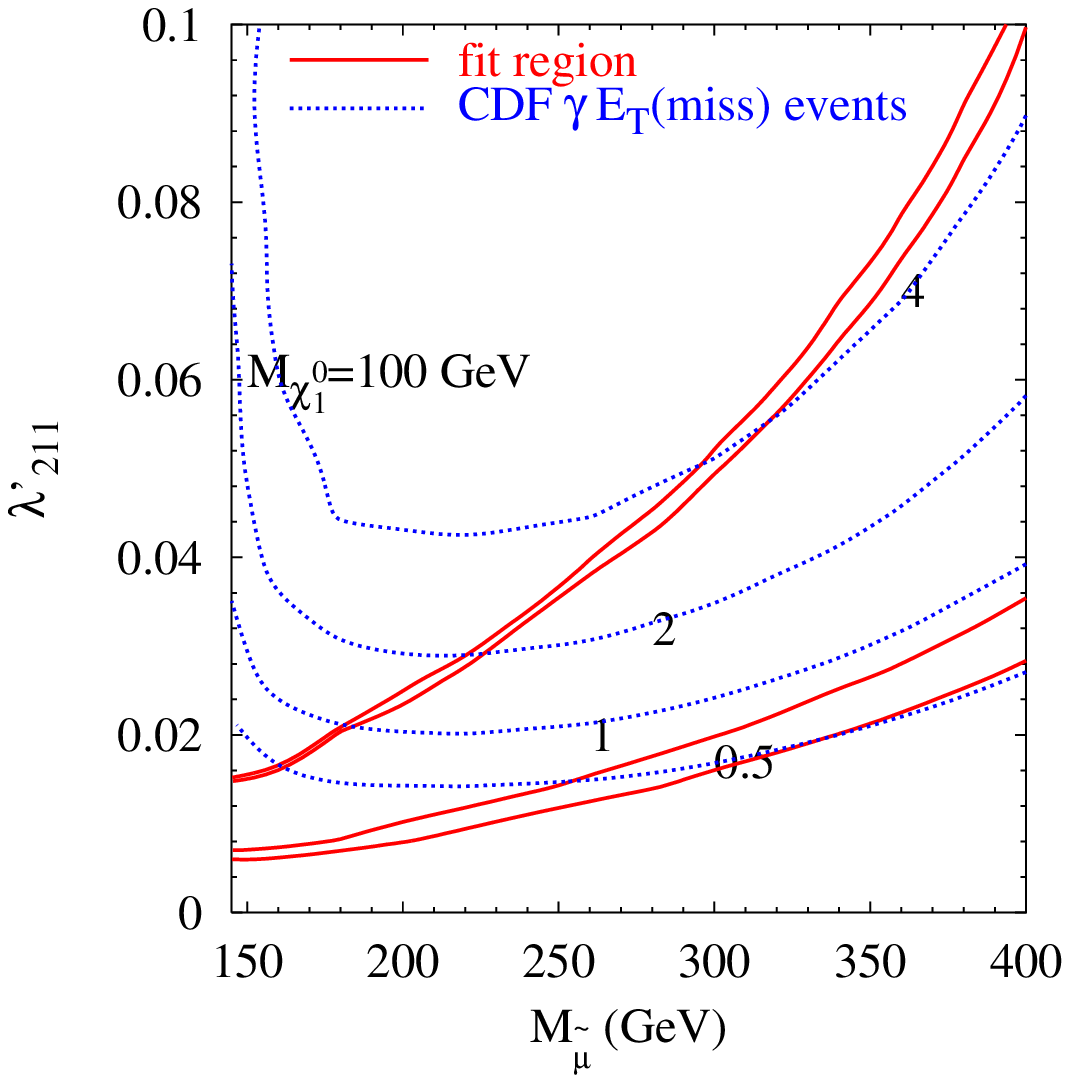}
\caption{Predicted number of excess events in (a) CDF $l \gamma \mett$,
  (b) D0 $l \gamma \mett$,
  and (c) CDF $\gamma \mett$  channels in the $\lambda'_{211}$-smuon
  mass plane, for $\tan \beta=10$, $m_{\tilde G}=10^{-3}$ eV and
  $M_{\chi_1^0}=100$ GeV. Labelled contours of equal numbers of signal events are shown as
  dashed curves.
  The 90$\%~ C.L.$ and 95$\%~C.L.$ combined-fit region is
  between the inner and outer pair of solid lines respectively. 
\label{lamsmu}
}
\end{figure}
Fig.~\ref{lamsmu}  
displays the $90$ and $95\%~C.L.$ fit regions as the area
between the solid lines in the $\lambda'_{211}, M_{\tilde \mu}$ plane. Here we
have chosen default values of $m_{\tilde G}=10^{-3}$ eV and $M_{\chi_1^0}=100$
GeV.
A significant amount of parameter space fits
the combined data. The
best-fit point is $\lambda'_{211}=0.114$,
$M_{\tilde \mu}=154$ GeV, with $\Delta \chi^2 = 6.90$. 
Fig.~\ref{lamsmu}a 
shows that we expect between 1-12
signal events in the CDF $l \gamma \mett$ channel. 1-8 events are 
expected at D0, and up to 4 signal events were predicted for the CDF
$\gamma \mett$ signature, depending on the parameters.

\begin{figure}%
\threegraphs{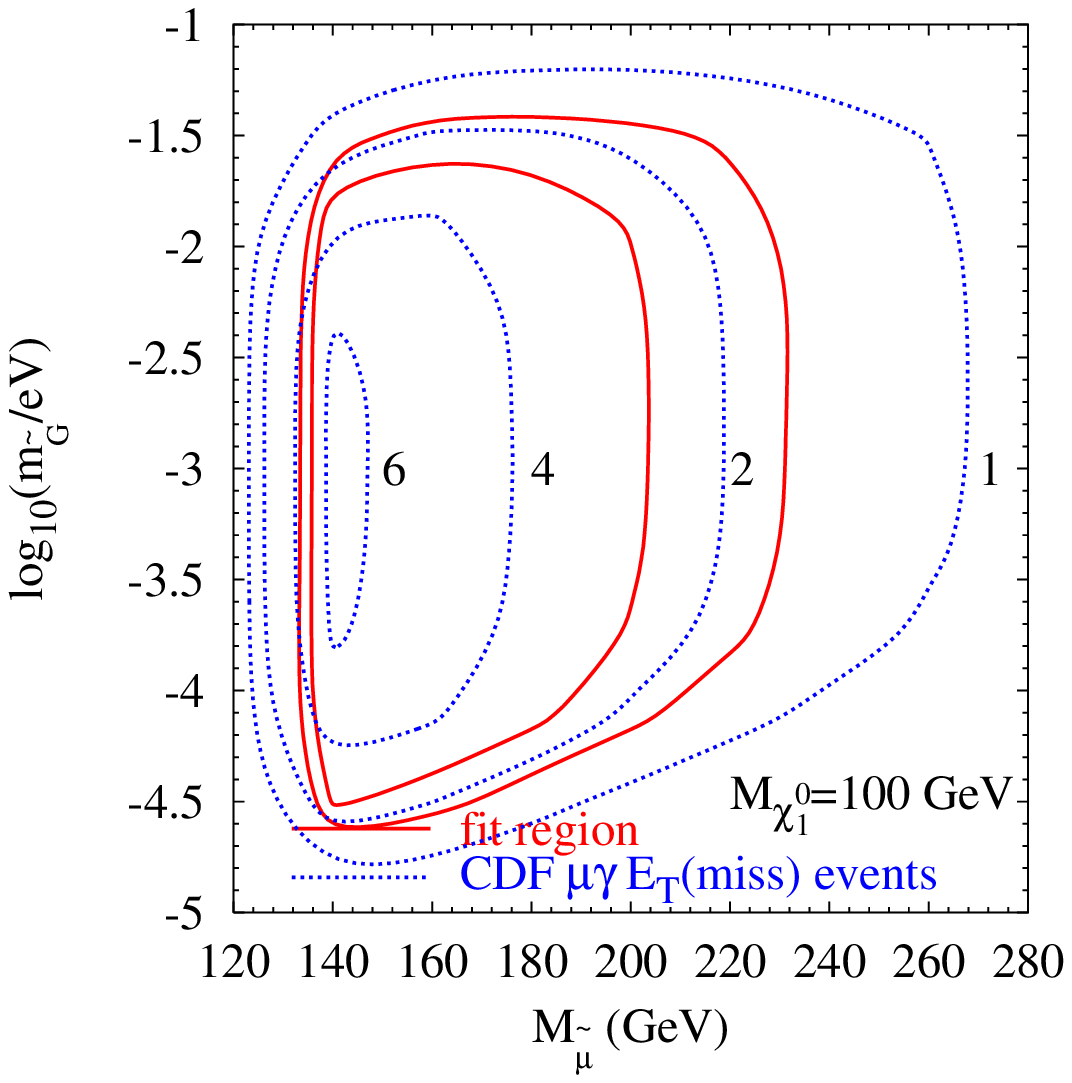}{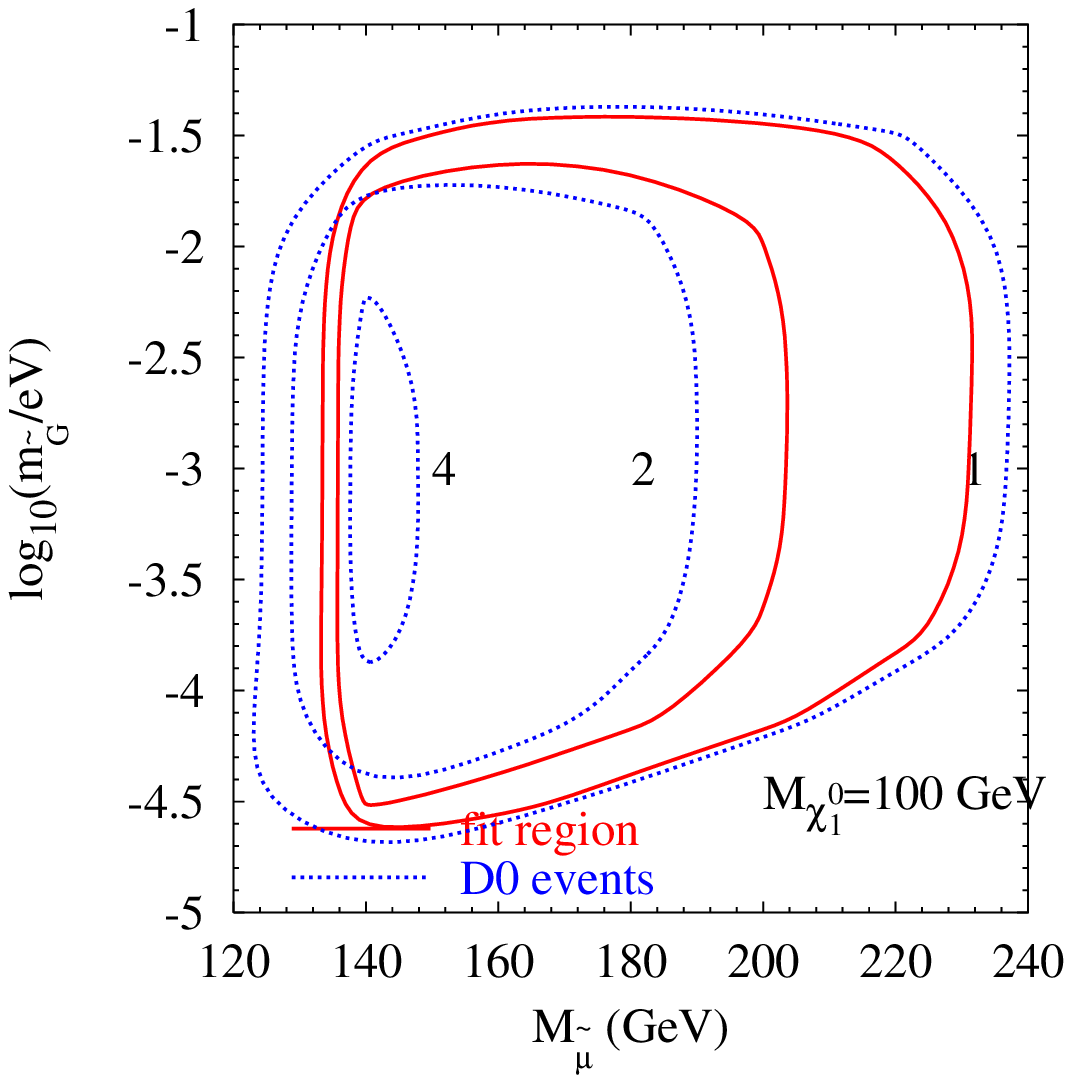}{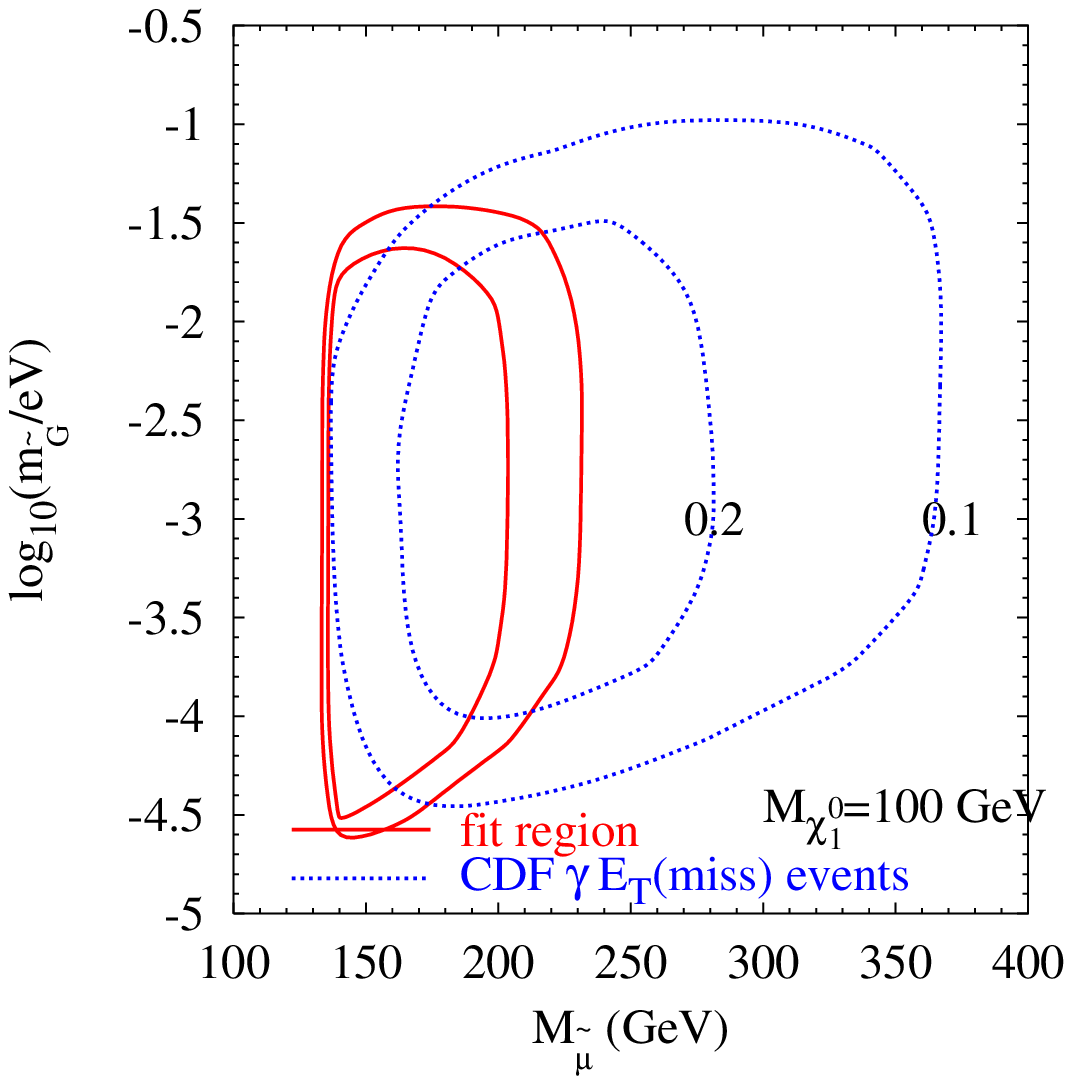}
\caption{Predicted number of excess events in (a) CDF $l \gamma \mett$,
  (b) D0 $l \gamma \mett$,
  and (c) CDF $\gamma \mett$  channels in the gravitino mass-smuon
  mass plane, for $\tan \beta=10$, $M_{\chi_1^0}=100$ GeV and
  $\lambda'_{211}=0.01$. Labelled contours of equal numbers of signal events
  are shown as dashed curves.
  The 90$\%~ C.L.$ and 95$\%~C.L.$ combined-fit regions are
  depicted by solid lines. }
\label{mgsmu}
\end{figure}
Fig.~\ref{mgsmu}  
displays the $90$ and $95\%~C.L.$ fit regions as the area
enclosed by the solid lines in the $m_{\tilde G}, M_{\tilde \mu}$ plane. Here
we 
have chosen default values of $\lambda'_{211}=0.01$ and $M_{\chi_1^0}=100$ GeV.
The ranges $M_{\tilde \mu}=130-230$ GeV, $m_{\tilde G}=10^{-4.5}-10^{-1.5}$ eV
provide a reasonable combined fit.
The
best-fit point is $m_{\tilde G}=10^{-3.1}$ eV,
$M_{\tilde \mu}=142$ GeV, with $\Delta \chi^2 = 6.76$. 
Fig.~\ref{mgsmu}a 
shows that we expect between 1-6
signal events in the CDF $l \gamma \mett$ channel. 0-4 events are 
expected at D0, and up to 0.2 signal events were predicted for the CDF
$\gamma \mett$ signature, depending on the parameters. 

\section{Conclusions}

We have provided combined fits for a supersymmetric model that explains the
$l \gamma \mett$ CDF Run I excess in events, which was at the 2.7$\sigma$
level~\cite{CDF}. We have used the Run I $\gamma 
\mett$ data recently presented by CDF, as well as anomalous trilinear gauge
boson coupling data from D0. Constraints upon various hyper planes in the
gravitino, smuon, neutralino and R-parity violating coupling space have been 
displayed. In totality, the signal rates predicted by our model for the three
analyses fit the data well, best fit points corresponding to a $\Delta
\chi^2=6.9$ fit compared to the Standard Model, for one degree of freedom.

Unfortunately, background rates in the D0 anomalous trilinear gauge
boson coupling data are too high for it to be very sensitive to the 
predicted signal rate. We note, however, that another analysis on existing D0
Run I data with cuts optimised to test a $\mu \gamma \mett$ excess would 
provide a good test of our scenario. 

The CDF$\gamma \mett$ channel suffers from a high $E_T(\gamma)>55$ cut because
of cosmic backgrounds, which
unfortunately also cuts most of the signal. It was shown in ref.~\cite{us},
that if the cut could be reduced to 25 GeV, a signal rate of several times
that in the CDF $l \gamma \mett$ channel is possible. This is an important
observation for Run II, where cosmic backgrounds could be cut by additional
timing information in the detector. We look forward to the analysis
of Run II data, which will be the final arbiter on $l \gamma \mett$ excess, 
as well as on our scenario.

\section*{Acknowledgements}
We would like to thank J. Berryhill, H.J. Frisch, S. Hagopian, D. Toback and
D. Wood for help and advice regarding experimental analyses and data. We would
also like to thank CERN, where much of this work was carried out.


\begin{thebibliography}{99}

\bibitem{CDF}
D. Acosta et al, Phys. Rev. D66 (2002) 012004, hep-ex/0110015;
D. Acosta et al, Phys. Rev. Lett. 89 (2002) 041802, hep-ex/0202044.

\bibitem{us}
B.C. Allanach, S. Lola and K. Sridhar, Phys. Rev. Lett. 89 (2002) 011801,
hep-ph/0111014; 
B.C. Allanach, S. Lola and K. Sridhar,
JHEP 0204 (2002) 002, hep-ph/0112321.

\bibitem{D0}
S. Abachi et al, Phys. Rev. Lett. 78 (1997) 3634, hep-ex/9612002.

\bibitem{CDF2}
D. Acosta et al, hep-ex/0205057;
T. Fahland, Ph. D. thesis (1997) Brown University, can be viewed at URL
\vspace{-0.5cm}
\begin{verbatim}
http://www-d0.fnal.gov/results/publications_talks/thesis/fahland/thesis.ps
\end{verbatim}

\bibitem{softsusy}
B.C. Allanach, Comput. Phys. Comm. 143 (2002) 305, hep-ph/0104145.

\bibitem{gmsb}
G. Giudice and R. Rattazzi, Phys. Rept. 322 (1999) 419 and references
therein.

\bibitem{baer}
H. Baer, M. Brhlik, C. Chen and X. Tata, Phys. Rev. D55 (1997) 4463.

\bibitem{kane} 
S. Ambrosanio, G. L. Kane, G. D. Kribs, S. P. Martin, and S. Mrenna,
Phys. Rev. Lett. 76 (1996) 3498;
G. L. Kane and S. Mrenna, Phys. Rev. Lett. 77 (1996) 3502;
S. Ambrosanio, G. L. Kane, G. D. Kribs, S. P. Martin, and S. Mrenna, 
Phys. Rev. D55 (1997) 1372.

\bibitem{abe}
F. Abe et al, Phys. Rev. D59 (1999) 092002.

\bibitem{witten}
E. Witten, hep-ph/0201018 and
talk at SUSY 2002, can be viewed at URL
\vspace{-0.5cm}
\begin{verbatim}http://www.desy.de/~susy02/pl.6/witten.pdf\end{verbatim}

\bibitem{bgh}
V. Barger, G. F. Giudice and T. Han, Phys. Rev.
D 40 (1989) 2987.

\bibitem{rparrev}
B. C. Allanach, A. Dedes, and H. K. Dreiner,
Phys. Rev. {D60}, 075014 (1999), hep-ph/9906209;
H. Dreiner, `Perspectives on Supersymmetry', Ed.\ by 
G.L. Kane, World Scientific.

\bibitem{cdfjets}
F. Abe et al, Phys. Rev. D55 (1997) 5263.

\bibitem{isajet}
F. E. Paige, S. D. Protopescu, H. Baer and X. Tata,
hep-ph/9810440.

\bibitem{herwig}
G. Corcella et al, hep-ph/0201201;
G. Marchesini, B. R. Webber, G. Abbiendi, I. G. Knowles, M. H. Seymour and
L. Stanco, JHEP 01 (2001) 010 hep-ph/0011363; {\em ibid.} hep-ph/0107071.
``HERWIG: A Monte Carlo event generator for simulating hadron emission
reactions with interfering gluons. Version 5.1 - April 1991'',
{Comput. Phys.Commun.}  {67} (1992) 465.

\bibitem{minuit}
F. James and M. Roos, Comput. Phys. Commun. 10 (1975) 343.

\bibitem{lep2}
LEPSUSYWG, ALEPH, DELPHI, L3 and OPAL experiments, note
LEPSUSYWG/02-07.1
\vspace{-0.5cm}
\begin{verbatim}
http://lepsusy.web.cern.ch/lepsusy/Welcome.html
\end{verbatim}
\end{thebibliography}
\end{document}